\documentclass[useAMS,usenatbib]{mn2e}

\usepackage[dvips]{graphicx}  
\usepackage{color}
\usepackage{aas_macros}  
\usepackage{amssymb}

\newcommand{\Rsolar}{\mbox{\,$\rm R_{\odot}$}}        

\title[{\sl K2} observations of EQ\,Psc]{{\sl K2} observations of the pulsating 
subdwarf B star  EQ\,Piscium: an sdB+dM binary } 
\author[C. S. Jeffery \& G. Ramsay]{C. S. Jeffery$^{1,2}$ and G. Ramsay$^1$  \\
$^{1}$Armagh Observatory, College Hill, Armagh BT61 9DG, UK\\
$^{2}$School of Physics, Trinity College Dublin, College Green, Dublin 2, Ireland}
\begin{document}


\pagerange{\pageref{firstpage}--\pageref{lastpage}} \pubyear{2014}

\maketitle

\label{firstpage}

\begin{abstract}
 {\sl K2},  the two-wheel mission of the  {\sl Kepler} space telescope, 
observed the pulsating subdwarf B star EQ\,Psc during engineering tests in 2014 February.  
In addition to a rich spectrum of g-mode
pulsation frequencies, the observations demonstrate a 
light variation with a period of 19.2\,h and a full amplitude of 2\%.  
We suggest that this is due to reflection from a cool companion, making EQ\,Psc the longest-period
member of some 30 binaries comprising a hot subdwarf and a cool dwarf
companion (sdB+dM), and hence useful for exploring the common-envelope
ejection mechanism in low-mass binaries.
\end{abstract}


\section{Introduction}

The hot subdwarf EQ\,Psc (=PB\,5450) was one of a number of stars
discovered to show long-period pulsations likely to be due to
opacity-driven gravity-modes \citep{green03}. Both gravity and
pressure-mode pulsations are found amongst a significant fraction of
hot subdwarfs, and offer the possibility to study the stellar interior
using the methods of asteroseismology. During the first two-years of
operation of the {\sl Kepler} spacecraft, some twenty pulsating hot
subdwarfs were identified in the {\sl Kepler} field and observed for
one or more quarters. The majority of these stars are primarily cool
V1093 Her (g-mode) variables, one is a hybrid DW Lyn (p-mode + g-mode)
variable and one is a low-amplitude V361 Her (p-mode) variable,
although very-low amplitude p-modes are detected in a substantial
fraction of the {\sl Kepler} V1093 Her variables
\citep{kepler10.2,kepler10.3}.

In addition to pulsation-driven light-variations, subdwarf B stars in
binary systems can also show light variations caused by eclipses {
and/or by reflection from} a companion (HW\,Vir variables) \citep{menzies86},
ellipsoidal deformation in a very short-period binary ({\it e.g.}
KPD\,1930+2752) \citep{billeres00} and Doppler beaming ({\it e.g.}
2M\,1938+4603) \citep{barlow12}.  Such phenomena are vital tools for
interpreting the fundamental properties, internal structure and
evolutionary origin of hot subdwarfs which, by several means, have
become nearly-naked core-helium-burning stars \citep{heber86,han02}.

Because of the rarity of hot subdwarfs and the significance of light
variability for understanding them, it is natural that any such star
which falls within a {\sl Kepler} pointing should, if possible, be
observed. EQ\,PSc happened to fall in a field used for a nine-day
engineering test at the start of 2014 to verify the stability of
operations with two reaction wheels ({\sl K2}) \citep{howell14}. In this
letter, we report observations which show a 19\,h periodic variation,
which we interpret as being due to reflection from a cool stellar
companion, as well as a rich g-mode pulsation spectrum.

\begin{figure}
	\centering
		\includegraphics[width=0.47\textwidth,angle=0]{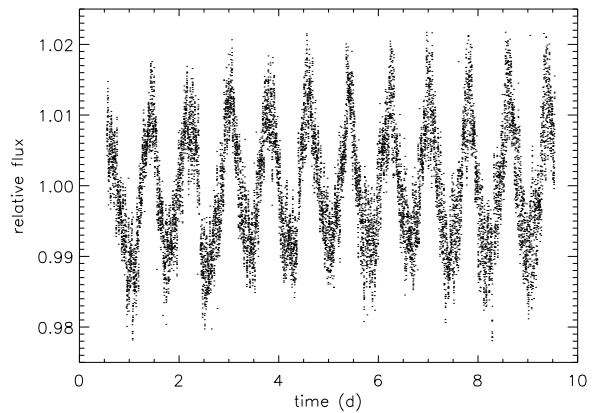}
\caption{The {\sl K2} light curve for EQ\,Psc in relative flux units. Times in all figures 
       are given as MJD - 56692.}
	\label{f:lc}
\end{figure}

\begin{table}
\label{t:freq}
\caption{Frequencies identified in the {\sl K2} power spectrum of EQ\,PSc.}
\begin{center}
\begin{tabular}{lrrrr}
\hline
 & $f\, {\rm (d^{-1})}$ & P (s) & $a$ (ppt) & $\phi$~~~~~ \\
\hline
$f_{\rm orb}$  &  1.2478 & 69242 & 10.218(107)& 0.494(02)  \\
$2f_{\rm orb}$ &  2.4956 & 34621 & 1.300(108)& 1.240(14)  \\
\hline
$f_{\rm L+}$    &  8.0831 & 10697 & 0.569(110)& 0.098(31)  \\
$f_{\rm L-}$    &  8.2256 & 10512 & 0.611(110)& 0.089(29)  \\
$f_{\rm K}$    & 17.0468 &  5085.4 & 0.986(108)& 0.373(17)  \\
$f_{\rm J}$    & 17.8751 &  4851.4 & 1.506(108)& 1.235(11)  \\
$f_{\rm I}$    & 24.6959 &  3523.3 & 0.375(108)& 0.334(46)  \\
$f_{\rm H}$    & 26.9578 &  3232.0 & 1.073(110)& 0.898(16)  \\
$f_{\rm H-}$    & 27.0933 &  3216.1 & 0.321(110)& 0.423(54)  \\
$f_{\rm G}$    & 29.2696 &  2981.1 & 0.943(108)& 0.286(19)  \\
$f_{\rm F}$    & 33.8270 &  2587.3 & 0.360(108)& 0.912(48)  \\
$f_{\rm E+}$   & 35.2728 &  2484.8 & 0.723(111)& 0.177(24)  \\
$f_{\rm E}$    & 35.3638 &  2478.5 & 1.601(111)& 0.869(11)  \\
$f_{\rm D}$    & 39.9094 &  2204.8 & 0.451(108)& 0.393(38)  \\
$f_{\rm C}$    & 42.8191 &  2060.6 & 2.322(108)& 0.634(07)  \\
$f_{\rm B+}$   & 46.6886 &  1897.2 & 6.573(381)& 0.557(09)  \\
$f_{\rm B-}$    & 46.7061 &  1896.6 & 6.698(381)& 0.124(09)  \\
$f_{\rm A}$    & 73.6477 &  1246.8 & 0.424(108)& -0.237(41)  \\
\hline
\end{tabular}
\end{center}
\end{table}

\begin{figure*}
	\centering
		\includegraphics[width=0.9\textwidth,angle=0]{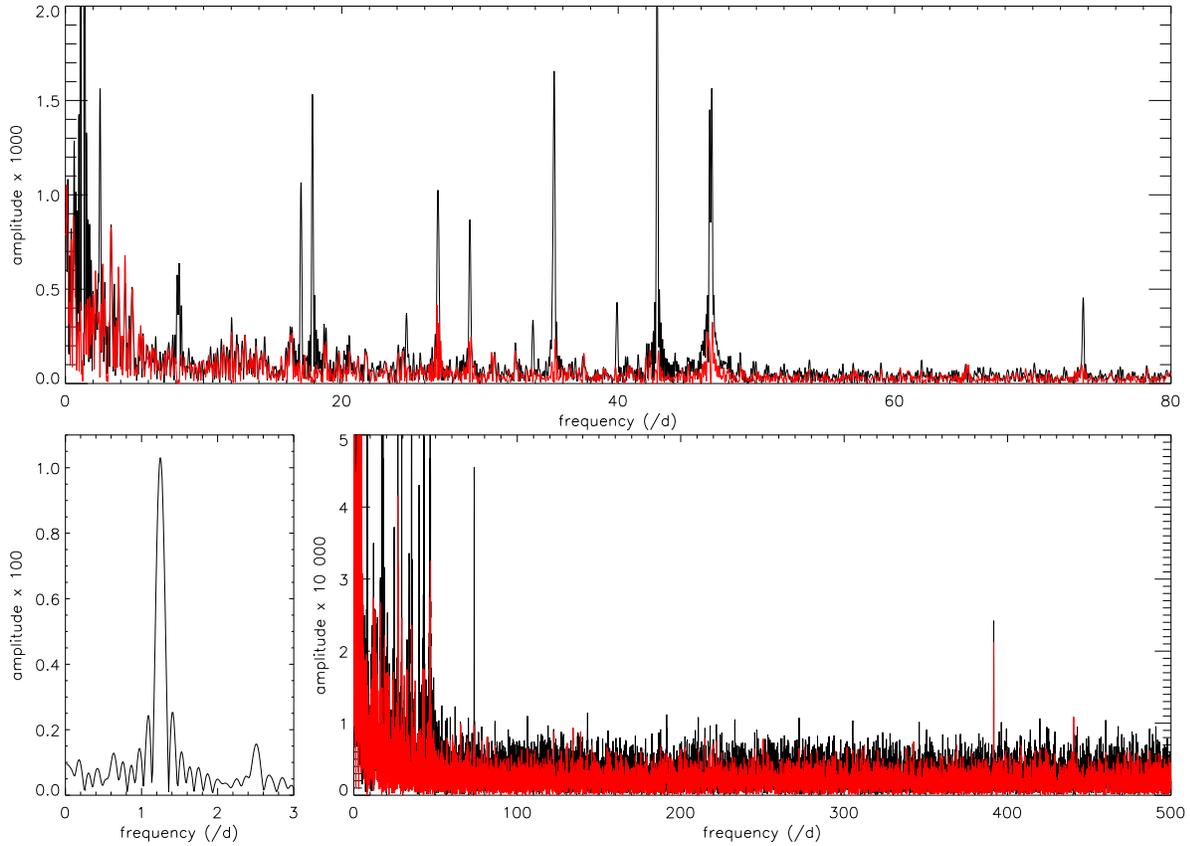}
\caption{Amplitude spectra obtained from the {\sl K2} light curve of
  EQ\,Psc. In all panels, black represents the power spectrum of the
  data shown in Fig.~\ref{f:lc}, and red represents the power spectrum
  of the data pre-whitened by the frequencies and amplitudes shown in
  Table~\ref{t:freq}. }
	\label{f:power}
\end{figure*}

\section[]{{\sl K2} observations and data reduction}

The detector on board {\sl Kepler} is a shutterless photometer using 
6\,s integrations and a 0.5\,s readout. The observations of EQ\,Psc
were made in short cadence (SC) mode, where 9 integrations are summed for
an effective 58.8 sec exposure.  Observations were carried out in
engineering mode from {MJD 56692.5411 to 56701.5306 } (2014 Feb 4th to Feb
13). The coverage was therefore 8.9 days in duration. During this time
interval there were frequent corrections to the spacecraft pointing, 
with one significant shift occuring on MJD=56694.86 (or
2.3 days into the time series).

A 50$\times$50 pixel array is downloaded from the satellite for each
target.  To extract a light curve of EQ Psc we used the {\tt PyKe}
software (Still \& Barclay
2012){\footnote{http://keplergo.arc.nasa.gov/PyKE.shtml}} which was
developed for the {\sl Kepler} mission by the Guest Observer Office.
We experimented by extracting data from a series of different
combinations of pixels. We found that a mask centered on EQ Psc
consisting of 110 pixels gave the optimal results. If a smaller number
of pixels are used we find that there are small discontinuities
present in the light curve which are the result of small shifts in the
position of the stellar profile over the CCD.  We also experimented
with subtracting the background (which increased in a nearly linear
fashion over the course of the observations) in different ways. We
found that using the median value of each time point to represent the
background gave the best results. Finally we removed time points which
were not flagged {\tt `SAP\_QUALITY==0'} (for instance during times of
enhanced solar activity). The extracted and reduced light curve is
shown in Fig.~\ref{f:lc}

\begin{figure*}
	\centering
		\includegraphics[width=0.9\textwidth,angle=0]{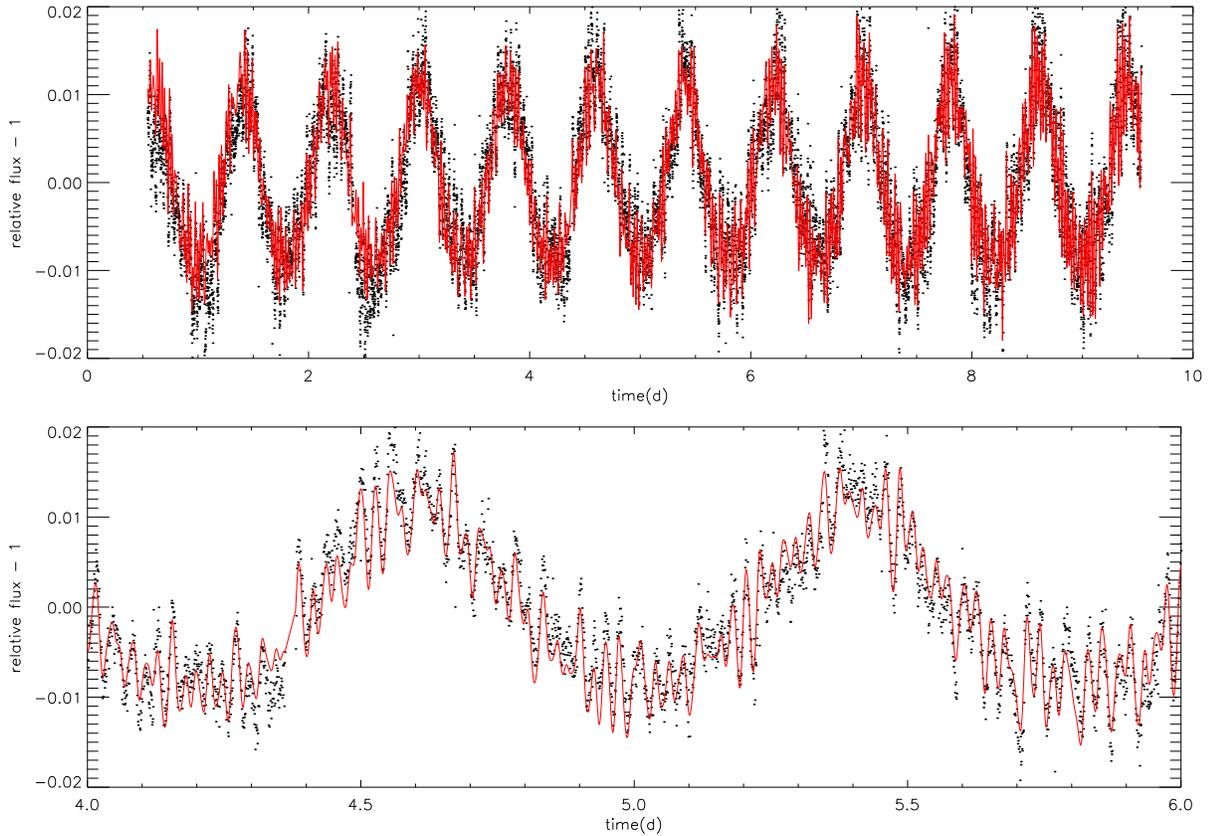}
\caption{The adopted fit to the {\sl K2} light curve for EQ\,Psc shown in
  Fig.~\ref{f:lc}.  Black dots are the {\sl K2} observations. The solid red
  curve shows the solution represented in Table ~\ref{t:freq}.  }
	\label{f:fit}
\end{figure*}

\section[]{Light curve and frequency analysis}

The {\sl K2} light curve for EQ\,PSc shows a strong modulation with an
amplitude of about two per cent (0.022 mag) and a period close to 0.8 d
(Fig.~\ref{f:lc}). Closer inspection shows additional variability on a
timescale of half an hour with an amplitude of around one per cent.
 
The periodic content of the light curve was investigated using a
classical power spectrum analysis. An idea of the window function can
be obtained from the lower-left panel in Fig.~\ref{f:power}.  For a
continuous run lasting 9\,d, the nominal frequency resolution is
$0.22\,{\rm d^{-1}}$.
Peaks in the power spectrum were identified in order of descending
power, amplitudes and phases associated with each peak were measured
using a multi-sine fit to the light curve, the light curve was
pre-whitened by this fit, a new power spectrum was computed, the
next-highest peaks were identified and added to the frequency table,
and the cycle was iterated.

A provisional list of frequencies in cycles per day, periods in
seconds, semi-amplitudes in parts per thousand (ppt\footnote{Also
  referred to as {\it milli-modulation intensity} units, or mmi.}),
and phases representing the {\sl K2} lightcurve of EQ\,Psc is
presented in Table~\ref{t:freq}. {Phases refer to a zero-point at MJD
  56692.0 and are given in cycles.  Errors are shown in parentheses.}
The multi-sine fit constructed from the data in this table is
illustrated in Fig.~\ref{f:fit}. Frequencies have been labelled in
alphabetical order of increasing period.

Several low-frequency peaks were excluded from this analysis; with the
exception of the dominant peak at $1.248\,{\rm d^{-1}}$, its first
harmonic and their aliases, the power spectrum is dominated by red
noise at $f<5\,{\rm d^{-1}}$.  At $f>100\,{\rm d^{-1}}$, the power
spectrum has the characteristics of all {\sl Kepler} SC data
\citep{baran13}, including a peak at $\approx 390\,{\rm d^{-1}}$ and a
picket fence of low-amplitude peaks just visible in the lower-right panel
of Fig.~\ref{f:power}.  None of these were considered to have a
stellar origin. The highest-frequency signal $f_{\rm A}$ ($P_{\rm
  A}\approx21$\,m) is isolated from than the main group of g-mode
frequencies. It may be a {\sl K2} artefact. There is a variable artefact at
around $31 - 32\,{\rm d^{-1}}$ \citep{baran13}, but this is well clear
of $f_{\rm F}$. $f_{\rm K}$ is close to a known broad feature at 
$16.98\,{\rm d^{-1}}$ in {\sl Kepler} SC data. 

There remain a number of significant peaks in the power spectrum after
subtracting the 18-frequency solution shown here. These are all
low-amplitude partners to much stronger peaks; examples include the
base of the doublet at $46.7\,{\rm d^{-1}}$ and peaks at 27.1 and
$29.3\,{\rm d^{-1}}$ Given the frequency resolution of our data, these
may not be real. This is demonstrated by the large amplitudes in the
fit to the $46.7\,{\rm d^{-1}}$ doublet, which do not reflect the
height of the peaks in the power spectrum.  The clue is in the phases,
which are almost half a cycle different at the start of the run.
Since the beat period between these two frequencies is some 57 days,
the oscillations in the model are almost in anti-phase for the
duration of the observations. As for other V1093\,Her variables, it
will require a much longer observing run to fully resolve the g-mode
oscillations in this star.

In contrast to the sixteen or more pulsating sdB stars previously
observed with {\sl Kepler} \citep{kepler10.6,kepler10.7}, the
frequency resolution of the current data limits the information that
can be extracted for asteroseismic analysis. It would be useful, for
example, to identify series and multiplets with large and small period
spacings, respectively, but this is not possible here. 
Where two frequencies in Table~\ref{t:freq} 
are identified with a spacing less than the nominal
experimental resolution ($0.22\,{\rm d^{-1}}$), they have been marked
B--, B+, etc. If one component is much stronger than the
near-neighbour, the suffix is omitted from the strongest component.

Frequencies $f_{\rm B} - f_{\rm L}$, corresponding to periods in the
range 1800 - 11000\,s, are typical of periods commonly seen in
V1093\,Her variables \citep{kepler10.6,kepler10.7}.  \citet{green03}
show 3\,h of R-band photometry for EQ\,PSc, which demonstrates
variability on timescales of $\approx30$\, minutes. This corresponds
well with the doublet $f_{\rm B-,B+}$.  Despite the nominal
resolution, this peak is clearly resolved in our power spectrum
(Fig,~\ref{f:power}) with a separation $\Delta f = 0.018\,{\rm
  d^{-1}}$.   We conclude that frequencies 
$f_{\rm B} - f_{\rm L}$ are associated with g-mode pulsations 
in EQ\,Psc. 

\begin{figure}
	\centering
		\includegraphics[width=0.47\textwidth,angle=0]{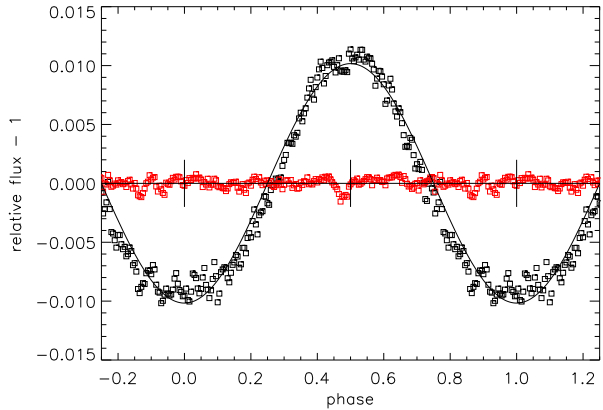}
\caption{The {\sl K2} light curve for EQ\,Psc binned ($\delta\phi=0.005$,
  black squares: relative flux - 1) and the residual light curve (red
  squares) folded on the orbital frequency. The first element of the
  fit (Table\,\ref{t:freq}) is shown as a solid line.  There is no
  evidence for an eclipse at either phase 0.0 (primary) or 0.5
  (secondary). }
	\label{f:orbit}
\end{figure}

\begin{figure}
	\centering
		\includegraphics[width=0.47\textwidth,angle=0]{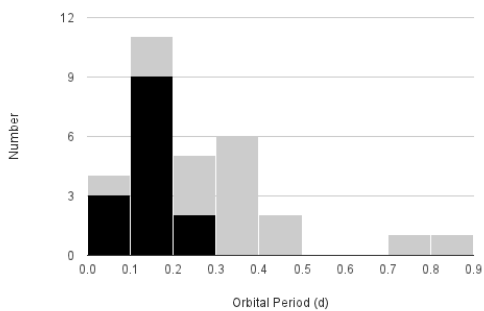}
\caption{The orbital-period distribution for known sdB+dM binaries, distinguishing 
eclipsing (HW\,Vir: black) and non-eclipsing (grey) systems.  }
	\label{f:periods}
\end{figure}

\section[]{Orbital reflection in a sdB+dM binary}

EP\,Psc shows a strong periodic signal at 0.801\,d (19.2\,h) with a
full {(peak-to-peak)} amplitude $>2\%$ of total light. The {\sl K2} data folded on this
period are shown in Fig.~\ref{f:orbit}.  The same data pre-whitened by
the Table\,\ref{t:freq} solution and filtered to remove low-frequency
(red) noise ($f < 3\,{\rm d^{-1}}$) are shown in the same figure.

The presence of an harmonic shows that this is not perfectly
sinusoidal. {Indeed, the roughly quarter-cycle phase difference between the
orbital harmonics flattens the minimum and sharpens the maximum. }
The lightcurve thus resembles those of hot subdwarf +
M-dwarf (sdB+dM) binaries in which the heated surface of the M-dwarf
facing the hot subdwarf reradiates (reflects) the incident
energy. Fourteen sdB+dM binaries with periods in the range 0.069 --
0.261\,d show such a reflection effect and also primary and secondary
eclipses: these are true HW\,Vir-type sytems. Fifteen sdB+dM binaries
with periods in the range 0.076 -- 0.75\,d show a reflection effect
alone with amplitudes ranging up to 0.1 magnitudes. The prototype is
XY\,Sex \citep{maxted02b}. {The longest-period system before
EQ\,Psc was JL\,82 (0.75\,d) \citep{koen09}. }

Since the 0.801\,d period is longer and the amplitude much larger than
observed in any known g-mode in a pulsating sdB star, pulsation would
seem to be a unsatisfactory explanation. Ellipsoidal deformation {of the hot star} 
due to rotation and tidal effects can be ruled out because the period is
too long (by a factor 10) to account for the amplitude
observed. Moreover, an ellipsoidal variation would have twice the
period measured, with minima of unequal depth.
 We conclude that reflection from an M-dwarf or {otherwise unseen companion} 
is the  most likely explanation for the 0.801\,d period in EQ\,Psc.  
Confirmation would make  EQ\,Psc the longest-period sdB+dM binary known to date.

In their paper reporting g-mode pulsations in EQ\,Psc, \citet{green03}
write "\ldots we did find three low-amplitude variables with periods
of several hours or more. Two are sdB stars exhibiting apparent
reflection effects, having appropriately phased sinusoidal light
curves the same length as their orbital periods ($\sim6$ and 12 hr)."
The 19.2\,h modulation would have been difficult to identify from the
3\,h photometry reported for EQ\,Psc, so its omission at that time is
understandable. During the preparation of this letter, the authors
have been made aware, however, that subsequent ground-based photometry
by Green et al.  does show evidence for a low-amplitude reflection effect 
(Green, {\O}stensen, priv. comm.).

Figure~\ref{f:orbit} shows no evidence for a primary eclipse, neither
in the total flux, nor in the residual flux. A dip just before the
expected position of secondary eclipes is consistent with being noise.
Geometric considerations alone imply that longer-period sdB+dM
binaries are less likely to eclipse, as indicated by the period
distributions of eclipsing and non-eclipsing systems
(Fig.~\ref{f:periods}\footnote{
{Periods for eclipsing (HW\,Vir) systems shown in Fig.~\ref{f:periods} 
were taken from: 
\citet{schaffenroth14} (SDSS\,J1622+4730: 0.0697\,d),
\citet{drechsel01} (HS\,0705+6700: 0.096\,d),
\citet{geier11} (SDSS\,J0820+0008: 0.096\,d),
\citet{kilkenny98} (NY\,Vir: 0.101\,d),
\citet{wils07} (NSVS\,14256825: 0.1104\,d),
\citet{ostensen07.wd} (HS\,2231+2441: 0.1106\,d),
\citet{menzies86} (HW\,Vir: 0.1168\,d),
\citet{barlow13} (EC\,10246-2707: 0.1185\,d),
\citet{polubek07} (OGLE BUL-SC16\,335: 0.1251\,d),
\citet{ostensen10} (2M\,1938+4603: 0.126\,d),
\citet{schaffenroth13} (ASAS\,102322-3737.0: 0.1393\,d),
\citet{for10} (2M\,1533+3759: 0.162\,d),
\citet{pribulla13} (FBS\,0747+725: 0.21\,d) and
\citet{kilkenny78} (AA\,Dor: 0.261\,d).
Periods for non-eclipsing (XY\,Sex) systems are from: 
\citet{maxted02} (XY\,Sex: 0.073\,d),
\citet{kupfer14} (UVEX\,J0328+5035: 0.11017\,d),
\citet{heber04} (HS\,2333+3927: 0.1718\,d),
\citet{morales03} (PG\,1329+159: 0.2497\,d),
\citet{ostensen13} (FBS\,0117+396: 0.252\,d),
\citet{for08} (2M\,1926+3720: 0.2923\,d),
\citet{geier14} (HS\,2043+0615: 0.3016\,d),
\citet{green04} (PG\,1438-029: 0.3358\,d),
\citet{latour14} (Feige\,48: 0.3438\,d),
\citet{kepler10.2} (KIC\,11179657: 0.3944\,d and KIC\,2991403: 0.4431\,d),
\citet{koen99b} (V1405\,Ori: 0.398\,d),
\citet{pablo11} (B4\,NGC6791: 0.3985\,d),
\citet{koen07b} (HE\,0230-4323: 0.4515\,d),
\citet{koen09} (JL\,82: 0.75\,d) 
and this paper (EQ\,Psc: 0.801\,d). }}
).

In crude terms, the amplitude of the reflection effect is a function of orbital 
separation $1/a$,  cool dwarf diameter $r_{\rm dM}$ 
and orbital inclination $i$.  Assuming that both components
are typical of other sdB+dM binaries. the orbital period indicates a
separation $a\approx3.0\pm0.2\Rsolar$, where the uncertainty indicates the
spread in the period-separation relation for shorter-period systems. 
Assuming representative radii of 0.17\,\Rsolar\ for both stars {\citep{almeida12}}, 
we deduce a maximum inclination $i\lesssim84^{\circ}$ to avoid eclipses.

The study of long-period sdB+dM binaries is important for 
stellar evolution theory and, in particular, for understanding the
common-envelope ejection process. sdB+dM binaries are believed to
originate in a system where an expanding red giant engulfs a low-mass
main-sequence companion shortly before core helium ignition \citep{han02}. 
The spiral-in of the  M dwarf leads to heating and ejection of the 
common-envelope surrounding the two stars. The final separation of
the components is indicative of the amount of orbital binding energy 
required to remove the red-giant envelope. In the case of EQ\,Psc, 
the long period may be the result of a wider-than-typical orbital separation, 
{and possibly linked with a higher-than-typical mass } for the M dwarf. 

\section{Conclusion}

Observations obtained during an engineering run with {\sl K2} in 2014
February have shown that  thruster-assisted, two wheel operation,
is sufficiently stable to obtain stellar photometry with high
precision.  The known pulsating subdwarf B star EQ\,Psc has been
demonstrated to show a 2\% light modulation with a period of
19.2\,h. It is argued that this is most likely due to reflection of
light from the hot subdwarf by a cool binary companion; an M dwarf is
likely.  The light curve also contains a rich spectrum of higher
frequency oscillations with periods in the range 1800 - 10\,000\,s. 
Most of these are likely to be associated with g-mode pulsations, as identified 
by \citet{green03}.  Spectroscopic studies of the binary orbit and of the 
hot star atmosphere are desirable.  Opportunities to explore the g-mode 
pulsation spectrum in more detail should be pursued.

\section*{Acknowledgments}

This paper includes data collected by the Kepler spacecraft using
thruster-assisted, two wheel operation. Funding for the Kepler spacecraft is provided by the
NASA Science Mission Directorate. The data were obtained through the
Mikulski Archive for Space Telescopes (MAST). STScI is operated by the
Association of Universities for Research in Astronomy, Inc., under
NASA contract NAS5-26555. Support for MAST for non-HST data is
provided by the NASA Office of Space Science via grant NNX09AF08G and
by other grants and contracts. This work made use of PyKE, a software
package for the reduction and analysis of Kepler data. This open
source software project is developed and distributed by the NASA
Kepler Guest Observer Office. We thank Tom Barclay, Martin Still and
Steve Howell for useful advice.  {We thank the referee for prompt and
constructive comments. }
Armagh Observatory is supported by
direct grant from the Northern Ireland Department of Culture, Arts and
Leisure.

\bibliographystyle{mn2e}
\bibliography{ehe}

\label{lastpage}

\end{document}